\documentclass[12pt, preprint]{aastex}
\newcommand{\cc}{\mbox{cm$^{-3}$}}

\def\m17{M~17}               % M 17    
\def\Htwo{H$_2$}               % H2    
\def\HtwoO{H$_2$O}             % H2O 
\def\thCO{$^{13}$CO}           % 13CO
\def\CeiO{C$^{18}$O}           % C18O
\def\Otwo{O$_2$}               % O2
\def\NtwoHp{N$_2$H$^+$}        % N2H+ 
\def\h{^{\rm h}}

\def\m{\ts {\rm m}}
\def\kms{\ts {\rm km\ts s$^{-1}$}}

\let\ts=\thinspace

\def\swash2o{$1_{10} - 1_{01}$}               % N2
\def\swas13co{J~$=~5~-~4$}
\def\isoh2o{2$_{12}-1_{01}$}               % N2
\def\h2oabun{$x$(o-H$_2$O)}
\def\oHtwoO{o-H$_2$O}               % H2O

\begin{document}
\setcounter{figure}{0}

\title{Sensitive Limits on the Water Abundance in Cold Low Mass Molecular Cores}

\author{Edwin A. Bergin}
\affil{Harvard-Smithsonian Center for Astrophysics, 60 Garden Street,
Cambridge, MA 02138} \email{ebergin@cfa.harvard.edu}

\author{Ronald L. Snell}
\affil{
Department of Astronomy,
University of Massachusetts, Amherst, MA 01003
} \email{snell@astro.umass.edu}

\begin{abstract}
We present SWAS observations of water vapor in two cold star-less clouds, B68 and Core D in 
$\rho$ Ophiuchus. Sensitive non-detections of the \swash2o\ transition of \oHtwoO\ are
reported for each source.  Both molecular cores have been previously examined by
detailed observations that have characterized the physical structure.  Using
these rather well defined physical properties and a Monte-Carlo radiation transfer
model we have removed one of the largest uncertainties from the abundance calculation
and set the lowest water abundance limit to date in cold low-mass
molecular cores.  These limits are $x$(\oHtwoO ) $< 3 \times 10^{-8}$ (relative to H$_2$) 
and $x$(\oHtwoO ) $< 8 \times 10^{-9}$ in B68 and $\rho$ Oph D, respectively.
Such low abundances confirm the general lack of ortho-water vapor in cold
(T $< 20$ K) cores.  Provided that the ortho/para ratio of water is not near
zero, these limits are well below theoretical predictions and appear to support 
the suggestion that most of the water in dense low-mass cores is frozen onto
the surfaces of cold dust grains.
\end{abstract}


\section{Introduction}

Water is an important molecule in the interstellar medium (ISM) because it
links chemistry in the ISM to comets and planetary systems and provides
crucial aid in constraining the chemistry of 
astrophysical systems \citep{bergin_imp}.   Recently
NASA's Submillimeter Wave Astronomy Satellite (SWAS) and
ESA's Infrared Space Observatory (ISO)
have provided the first unambiguous glimpse of water in the ISM.
Because the SWAS instrument is pre-tuned to the ground state rotational transition
of ortho-\HtwoO , it is sensitive primarily to water in the quiescent or coldest regions
of molecular clouds.  In contrast, ISO sampled higher energy transitions that consequently
probed warm
gas found in shocks or in close proximity to young embedded stars. 
One of the surprises from the analysis of SWAS results is that the derived water abundance
is unexpectedly low, with values \h2oabun\ $\sim 1 - 10 \times 10^{-9}$ (relative to
\Htwo ) found in a variety of regions (Snell et al. 2000a,b,c).
These abundances conflicted with theoretical expectations and several suggestions
have been provided to account for the discrepancy, with the principal solution 
for cold gas being the freeze-out of water onto grain surfaces 
\citep{bergin_imp, spaans_h2o, viti_h2o, charnley_swasiso}.   Moreover, ISO and SWAS detections of
water vapor in absorption
produced results in disagreement with the emission line analysis.   Absorption lines are more straightforward to
analyze and provide water abundances of \h2oabun\ $\sim 10^{-6}$,
in agreement with theory \citep{cernicharo_h2oabs, neufeld_h2oabs, moneti_h2oabs}.
These differences can be reconciled
provided that the freeze-out of water is greater in denser regions, which are
seen in emission, as opposed to the low density gas, with correspondingly longer
depletion timescales, seen in absorption \citep{bergin_imp}.

Nonetheless there are some well discussed uncertainties in the analysis of 
water emission lines detected by SWAS
(Snell et al. 2000a,b,c,).  These arise primarily from the fact that
only a small column of water is required to produce optically thick emission.
The situation is slightly simplified, because the
critical density for the \swash2o\ transition ($\sim 10^{8}$ \cc )
is higher than typical densities found in molecular cores.  Under this condition
every photon that is collisionally created eventually escapes and the water emission,
although optically thick, is effectively thin \citep{wannier_h2o, snell_omc1}.
This fact aids in constraining the abundance, but crucial
assumptions must be made regarding the source physical structure.  
Assuming that the \swash2o\ emission arises predominantly
from high density gas, simple single component models were constructed
(i.e. single density and temperature along the line of sight), resulting
in the above abundance estimates (Snell et al. 2000a,b,c).
However, uncertainties in the density are directly reflected in the abundance, and
a single density characterizing the entire line of sight is itself an approximation.  

More realistic physical structures were examined for a few 
molecular cores using a Monte-Carlo radiative transfer code \citep{ashby_cores}.  
This analysis supported the simple single component results.  However,
there are complications.  For instance in S140, a separate analysis
argues that a clumpy cloud model is more appropriate, although the resulting abundance
estimates are similar \citep{spaans_h2o}.  
%For $\rho$ Oph A the 
%presence of extended molecular emission (e.g. CS), and the expectation of
%strong temperature gradients needed to account for the infrared radiation field
%complicated the effort to construct a reliable physical model.  
Sources that offer the most promise to set good limits or determinations of
the water abundance 
are therefore ones that are largely 
isolated, and have well established physical properties.   In general, giant
molecular cloud cores, such as S140, do not fit this criteria 
as they have been suggested to have clumpy physical
structures which are difficult to model definitively.
On the other hand,
advances in the resolution and sensitivity of infrared (IR) and submillimeter continuum
arrays have begun to provide
a wealth of information on the physical structure of low mass cores.   
Studies using these instruments effectively apply similar techniques in
which azimuthally averaged dust column density
profiles are combined with assumptions regarding the geometry and a gas-to-dust ratio to 
determine the H$_2$ density profile.  

We present here SWAS observations of two molecular cores, B68 and $\rho$ Oph D, that have been
subject to these techniques and thus have rather reliable estimates of the 
source physical structure.    Using these physical models and SWAS observations
we have set the lowest limit to date on the water abundance in cold low mass 
cores.  The limits are below pure gas-phase chemical
predictions and support the assertion that most of the water is 
likely frozen onto grain surfaces in cold dense cores of molecular clouds.
 
\section{Observations and Results}

Between Feb. 25, 2001 to Mar. 8, 2001  SWAS observed B68 in the \swash2o\ transition
of \oHtwoO\ at 556.936 GHz for a total of 28 hours of on-source integration.  Similarly during 
Sept. 7-10, 2001 SWAS observed $\rho$ Oph~D for a total of 24.6 hrs (on-source).    System temperatures
were typically 2500 K with minimal scatter around that value.
The spacecraft
was used in nod mode, involving alternately nodding the spacecraft to an off-source 
position free of emission.   For B68 the elliptical  3\farcm3 $\times$ 4\farcm5 beam
was centered on $\alpha =$ 17:22:38.2, 
$\delta =$ $-$23:49:34 (J2000) and towards $\rho$ Oph~D, $\alpha =$ 16:28:30.4, 
$\delta =$ $-$24:18:29 (J2000).
All data were reduced with the standard SWAS
pipeline described by \citet{melnick_swas}.  
Towards both cores, along with \oHtwoO ,
SWAS simultaneously observed transitions
of [C I] ($^3$P$_1$ $\rightarrow$ $^3$P$_0$), \thCO\ (J = 5 $\rightarrow$ 4),
and \Otwo\ ($3_3 \rightarrow 1_2$).  In this work we primarily use the 
\oHtwoO\ and \thCO\ data, with a velocity resolution of 1.0 \kms , and a sampling
of 0.6 \kms .    All data are presented here on the T$_A^*$ scale and for 
subsequent analysis are scaled
by the main beam efficiency of 90\% \citep{melnick_swas}.

Figure 1 shows  spectra of the \swash2o\ transition of \oHtwoO\  and \thCO\
\swas13co\ taken towards
B68 and $\rho$ Oph~D.   
It is apparent that there are no water detections toward either source.
 In B68 the 3$\sigma$ upper limit is T$_A^* =$
36 mK and in $\rho$ Oph D the limit is 45 mK.   
The J $= 5 - 4$ transition of \thCO\ was not detected towards B68 with a similar
limit as \oHtwoO , confirming the cold (T $\le$ 10) nature of this source.    \thCO\ \swas13co\
is detected towards $\rho$ Oph D with an integrated intensity of 0.30 K km s$^{-1}$,
a peak antenna temperature of T$_A^* =$ 150 mK, and $\Delta$v $= 1.86$ km s$^{-1}$.

\section{Monte-Carlo Models of Water Emission}

To estimate the water abundance we draw upon previous determinations of the 
density structure derived from observations of dust in emission or 
absorption \citep{motte_oph, alves_b68, bacmann_dust}. 
%In general to derive the dust column density three techniques
%are used:  (1) dust emission maps at 450, 850, and 1300 $\mu$m 
%\citep[e.g.,][]{motte_oph, awb_ppiv,  shirley_dust},
%(2) infrared star counting techniques at J, H, and K bands (Lada et al. 1999; 
%Alves, Lada, \& Lada 2001), and (3) mapping the infrared absorption of cores
%against the diffuse mid--infrared background (Bacmann et al. 2000).   
%Each of these techniques suffer from various complications, but have greatly
%enhanced our understanding of the structure of low mass cores.
Towards both clouds, the SWAS beam encompasses the entire area seen in
transitions of molecules such as \CeiO , CS, and \NtwoHp .
Higher spatial resolution observations of molecular cores in these other tracers
are able to determine
the extent of emission, which along with the radial 
density profile with radius,  permits a more accurate analysis
of abundances
\citep{jessop_dep, tafalla_dep, bergin_b68, hotzel_b68}.
Because of the large beamsize our initial procedure is to assume 
a simple model with an abundance of \oHtwoO\ that is constant with cloud depth.
However, chemical models of centrally condensed objects 
predict that lower abundances should exist in the dense central
regions where molecules freeze onto grain surfaces more frequently
\citep{rawlings_dep}. 
% in the
%low density outer layers the gas-grain collision timescales are longer and
%higher abundances are predicted 
To account for this possibility we have also examined whether a 
water abundance profile predicted by theory is in conflict
with our observations.  This model is applied to B68 which has
a large amount of additional observational constraints.  
%We note that SWAS, with its larger beam, is insensitive to
%structures smaller than 4$'$ and thus SWAS data alone 
%cannot discriminate between simple constant abundance models and models
%with complex abundance structure.

The water abundance profile, 
along with the adopted density, temperature, and velocity width
profiles, is used as input into the spherical one-dimensional
Monte-Carlo radiative transfer code discussed by 
\citet{ashby_profile}.\footnote{Rates of excitation of water with 
ortho-H$_2$ are often an order of magnitude
greater than rates for lower energy para-H$_2$ \citep{phillips_h2oxs}.  
For our calculations of cold clouds we have assumed 
an ortho/para-H$_2$ ratio of 0.1 and extrapolated the rates down
to 10 K.  If the ratio is lower our results will be essentially
unchanged; a higher ratio would result in lower abundance limits.}
The adopted physical profiles for each source are motivated below, but
in each case we have assumed a static cloud.
The radiative transfer model determines the expected emission spectra, which is
compared to the observed 3$\sigma$ upper limit on the antenna temperature.   
The water abundance is
then iterated until the predicted emission matches the limit.   To account for 
oversampling in the SWAS spectrometer,
we have convolved the model data with a Gaussian  that has a width of 1.45 MHz.
Because of the weak IR continuum radiation field in both cores the effects of infrared
pumping are negligible.
%We include the effects of dust continuum on the water excitation in these models.
%But, as discussed in \citet{ashby_cores}, for these weak infrared sources the effects of infrared pumping are negligible.
%
\subsection{B68}

The B68 globule is located at a distance of 125 pc \citep{degeus_dist}.  \citet{alves_b68} examined this
isolated pre-stellar core using near-infrared extinction techniques.   The resulting 
extinction map was used to constrain the radial density profile,
which is well fit by a pressure confined, self-gravitating cloud
near equilibrium (Bonnor-Ebert sphere).  Recently,  
\citet[][hereafter BAHL02]{bergin_b68} and \citet{hotzel_b68} have examined this
cloud in the low-J mm-wave transitions of various molecules.  Both found 
that in the dense core center,
\CeiO\ molecules are systematically depleted onto grain surfaces.

We use the model described by BAHL02 
to estimate the water abundance. 
In this model the column density is constrained by the 
visual extinction data and the line width by \CeiO\ emission
(BAHL02).  
The assumed density profile is that derived from near-IR extinction measurements, while
the temperature structure is that for dust in a Bonnor-Ebert sphere 
derived by \citet{zucconi_tdust}.\footnote{However,
for the temperature, a uniform reduction of 2 K was required to match the multi-transitional
\NtwoHp\ observations of BAHL02.}  
The density and temperature structure are
provided in Figure~2.  This model assumes a static 
cloud, but includes contributions
from thermal and turbulent line widths with the latter increasing as a function of
radius.   

As a check on the assumptions regarding the cloud physical structure we first
examine the non-detection of \thCO\ \swas13co.
To model this emission we use the depleted \CeiO\ abundance profile
given by BAHL02 scaled by a factor of 7.8 (accounting for 
$^{12}$C/$^{13}$C = 75; $^{16}$O/$^{18}$O = 500).  Placing this into the
radiative transfer code  produces a peak temperature of  T$_A \sim 40$ mK,
which is close to the observed $3\sigma$ limit of 36 mK.    
Since the \thCO\ analysis does not significantly conflict with current limits, we then
use the iterative procedure outlined previously to estimate 
the 3$\sigma$ limit on the water abundance 
in B68, which is $x$(\oHtwoO ) $< 3 \times 10^{-8}$.    The resulting
emission spectra from the radiative transfer model are shown in Figure~1.\footnote{If 
the core radius is allowed to extend beyond 0.06 pc, 
as estimated by the near-IR analysis, 
then the abundance limit will be reduced.}

BAHL02 linked a radiative transfer code to a chemical model including 
the effects of molecular depletion to set limits on the 
freeze-out of \CeiO\ molecules.
Water is more tightly bound to grain
surfaces when compared to CO \citep{sa_binding}. Therefore it is more difficult
to remove and can be expected to be significantly depleted in the center, 
perhaps with larger depletions than CO.  
To see if current chemical theory is in
in conflict with our observational results we compare the predictions
of a gas-grain chemical model to the observations.   

To this end we use the gas-grain chemical model discussed in BAHL02.
The only difference in the chemistry is the 
additional inclusion of the grain surface formation of water 
via hydrogenation of oxygen as required by previous studies \citep{bergin_imp, viti_h2o,
charnley_swasiso}.   This model incorporates the cloud density and temperature profile 
shown in Figure~2.
We also include the observed cloud extinction 
profile which increases to A$_V$ = 17$^m$ at the core center.
The chemical model predicts the profile of water abundance with cloud radius which, 
along with the same physical profile, are placed as 
inputs to the radiative transfer calculations.  
Due to the inclusion of gas-grain interactions the chemical model
does not reach a steady state and the predicted abundance profile strongly 
varies with time.   Fortunately BAHL02 find that the observed C$^{18}$O and
N$_2$H$^{+}$ emission in B68 can be simultaneously reproduced by the
model at t  $\sim  7.6 \times 10^4$ yr (later times predict increasing depletion
and, in consequence, less C$^{18}$O emission than observed).
We therefore adopt this time
in our analysis to examine whether
the \HtwoO\ abundance profile predicted by the same model is
in conflict with SWAS observations. 
 
In Figure~3 we provide the predicted water vapor and ice abundances 
as a function of depth.   Also shown are the abundance of atomic oxygen and the
best fit constant abundance.  The chemical model predicts
that the water vapor abundance sharply declines
at the cloud edge due to photodissociation, rises to a peak at A$_V \sim 2^m$,
and then uniformly declines towards higher depths. 
As seen in Figure~3, nearly all of the water and oxygen
(in the form of H$_2$O)
is frozen onto dust grains.  
Placing the results of this model into the radiation transfer code we predict
a peak temperature in the \oHtwoO\ 557 GHz line of T$_A^*$ = 18 mK
below the 3$\sigma$ upper limit.  As long as the age is $> 10^{4}$ yr
our observations are consistent with this model.
Thus current predictions of water abundance profiles by chemical
models are not in conflict with observations.
%However, it must be stressed that the chemical model assumes that the chemistry
%is evolving while the density evolution is static.  Thus all quoted times
%represent minimum timescales as we do not account for any prior evolution
%at a lower density state.   
Given the potential of B68 as a template for testing chemical models 
it will be useful in the future
to re-examine these results in light of additional molecular observations.

\subsection{$\rho$ Oph D}

The $\rho$ Oph D molecular core is a star-less object
located in the Ophiuchus complex at a distance of 160 pc.
The core was detected in absorption against the galactic 
mid-IR background using ISOCAM \citep{bacmann_dust} and in 1.3 mm continuum emission \citep{motte_oph}.
These studies have provided good constraints on the core density profile, although
there are differences in the details.   Both agree that the density declines
as $\rho (r) \propto r^{-2}$ beyond $r$ $\sim 3300 - 4000$ AU; within this radius the
density is found to be constant, ranging from 
$n_{cst} = 3 \times 10^{5}$ \cc\ (Bacmann et al. 2000) to
$n_{cst} = 9 \times 10^{5}$ \cc\  (Motte et al. 1998).
Both studies derive a core radius of $\sim$ 13,000 AU (0.063 pc).  
However \citet{motte_oph} find a sharp edge to the South-West but, 
in the East-West direction, the dust emission merges with the ambient cloud. 
With these differences core mass estimates range from 2 -- 5 M$_{\odot}$.
In the following we examine both density distributions.

Unlike B68, where a detailed model exists, there is no 
information regarding the radial dependence of the velocity field or temperature in $\rho$ Oph D.
For temperature, the sub-millimeter dust continuum  survey of Ophiuchus by \citet{johnstone_oph} 
did not include core D, but for numerous other cores in the cloud they find T$_{dust}$ $\sim 10 - 30$ K, which
may be considered the expected range.   Given the lack of detailed information, we adopt the expression for the equilibrium dust temperature given by \citet{burton_pdr}, further assuming that T$_{dust} =$ T$_{gas}$.
This expression provides an estimate of the dust temperature depending on the 
extinction and of the local enhancement of the ultra-violet radiation field, G$_0$.    
For the  A$_V(r)$ profile we integrate the observed density 
profile with radius in a pencil beam along the line of sight. 
Because of the presence of early type stars the local radiation field is enhanced 
throughout the $\rho$ Oph cloud and detailed modeling of the 
far-infrared emission by  \citet{liseau_oph}  finds G$_0 = 20$ towards  core D. 

The procedure is to first examine how well the physical model(s) reproduce
the SWAS $^{13}$CO \swas13co\ emission and then apply the ``best'' model to the water
emission.   If we use the \citet{bacmann_dust} density profile, the above
temperature structure, and a constant turbulent width of 1.5 km s$^{-1}$ we can
match the observed \thCO\ emission provided $x$(\thCO ) $\sim 3 \times
10^{-6}$.  This abundance is higher than the expected range for 
undepleted gas.\footnote{Using the \CeiO\ abundance derived in the extended Ophiuchus cloud
by \citet{flw82} with plausible isotope ratios ($^{16}$O/$^{18}$O = 500; 
$^{12}$C/$^{13}$C = 45 -- 90) the \thCO\ abundance is expected 
to be 1 -- 2 $\times 10^{-6}$.}   Since this source is a cold pre-stellar
object we expect to see some evidence for \thCO\ depletion.  For instance in
B68, and several other similar pre-stellar objects,
large depletions are observed \citep{bergin_b68, jessop_dep, tafalla_dep}.

Adopting the denser profile of \citet{motte_oph} we can reproduce observed
\thCO\ emission at a lower abundance of $\sim$ 4 $\times 10^{-7}$. 
The predicted emission spectra from this model is shown superposed
on the observational data in Figure~1.   
For reproducing the water emission, the \citet{motte_oph} model is favored due to
the anticipation of CO depletion. 
Adopting this model for the  \swash2o\  observations we derive 
$x$(\oHtwoO ) $< 8 \times 10^{-9}$, relative to \Htwo (3$\sigma$). 
The model that matches the 3$\sigma$ limit is also provided in Figure~1.
If we use the Bacmann model the abundance limit 
is $x$(\oHtwoO ) $\lesssim 1 \times 10^{-7}$.
 
\section{Conclusions}

We have derived upper limits to the water abundance in two clouds with well
described physical properties.   These abundance limits are, 
$x$(\oHtwoO ) $< 8 \times 10^{-9}$ for
core D in $\rho$ Oph and $x$(\oHtwoO ) $< 3 \times 10^{-8}$ in B68.
These limits are below those previously set for a cold star-less object by \citet{snell_h2o}
(TMC-1: \h2oabun\ $< 7 \times 10^{-8}$).  In this Letter two cores have 
been subject to a more detailed and careful study which essentially 
confirms the general lack of ortho-water vapor in cold (T $< 20$ K) molecular cores.     
Provided  the ortho/para ratio of water is $>$ 0.03 then these results are in agreement 
with the assertion that the water abundance in low mass objects is well below the 
predictions of pure gas phase chemistry.  In the case of
B68, these observations are compared to theoretical predictions of
a gas-grain chemical model, which have been directly placed into the radiative transfer
calculations.  
We find that these model predictions are not in conflict with observations. 
However, due to the large SWAS beam, which encompasses the entire extent of the observed
molecular emission seen in other tracers, we cannot discriminate between
simple constant abundance models and those with complex abundance 
structure constrained by theory.
Given the wide-spread 
molecular depletion found in B68 and similar cores \citep{bergin_b68, hotzel_b68, 
tafalla_dep},  it is likely that most of the water in low mass dense cores 
is frozen on the surfaces of cold dust grains.

\acknowledgements

We acknowledge Gary Melnick for a thorough reading of the manuscript.
E.A.B. is grateful for the help and collaboration with Charlie Lada,
Jo\~ao Alves, and Tracy Huard which greatly aided the B68 analysis.
For these data we are grateful to entire SWAS team and acknowledge
support from NASA's SWAS Grant NAS5-30702.

\begin{center}
\begin{figure*}
\plotone{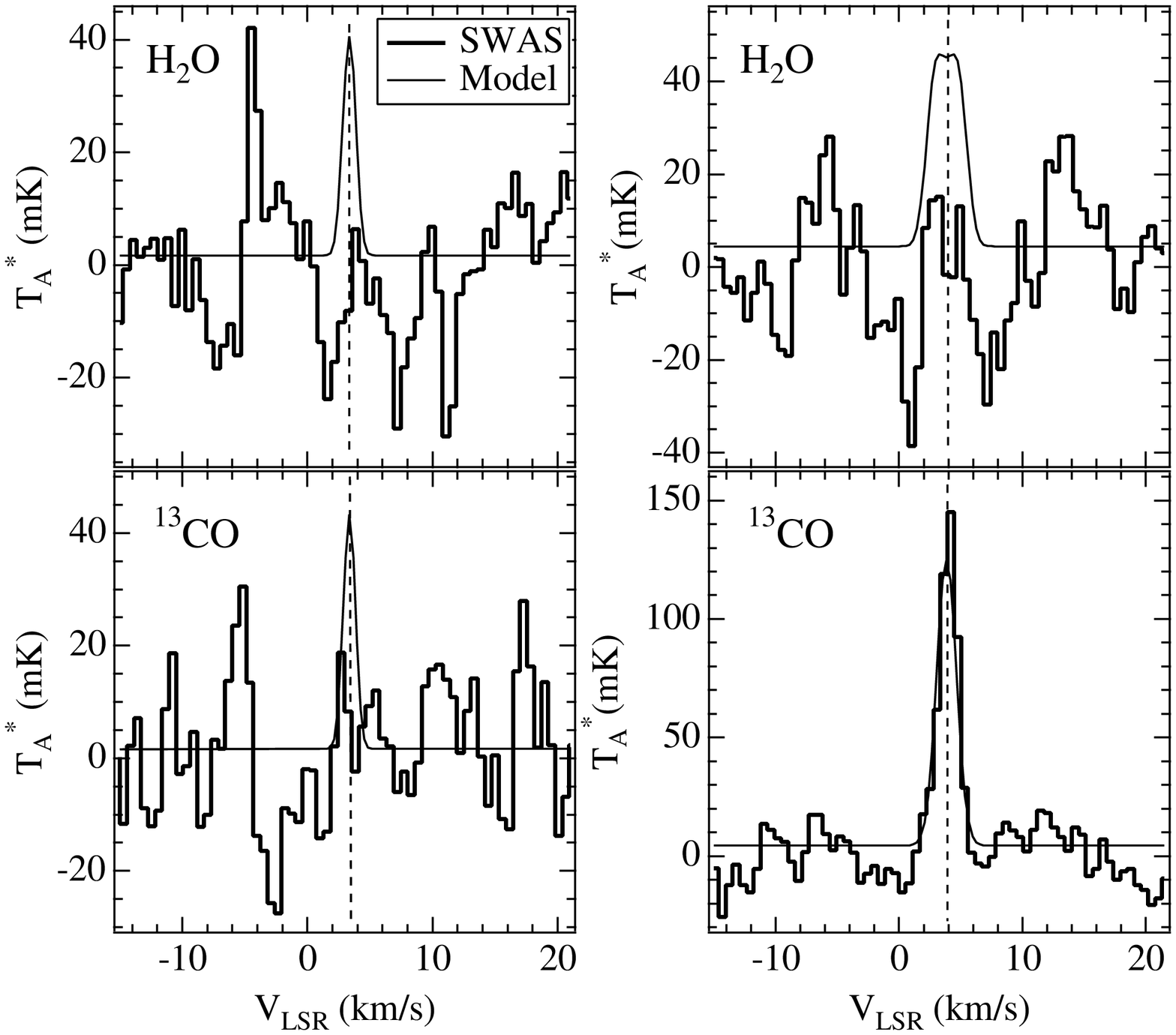}
\caption{
Spectra of the $1_{10} - 1_{01}$ transition of o-H$_2$O and $^{13}$CO J $= 5 - 4$
towards the B68 dark cloud (left) and $\rho$ Oph core D (right) shown as a
thick solid line.  The thin solid line is the result of the excitation model
described in \S3.1 and \S3.2, while the dashed line denotes the systemic velocity.
The model is shifted to the proper source velocity.  
$^{13}$CO J $= 5 - 4$ is detected in $\rho$ Oph D, but all other observations 
are non-detections.   In the figure a linear baseline has been subtracted from
the observational data.  The model spectra are shown with continuum included
which, due to the small level, has not been subtracted.
}
\end{figure*}
\end{center}

\begin{figure*}
\plotone{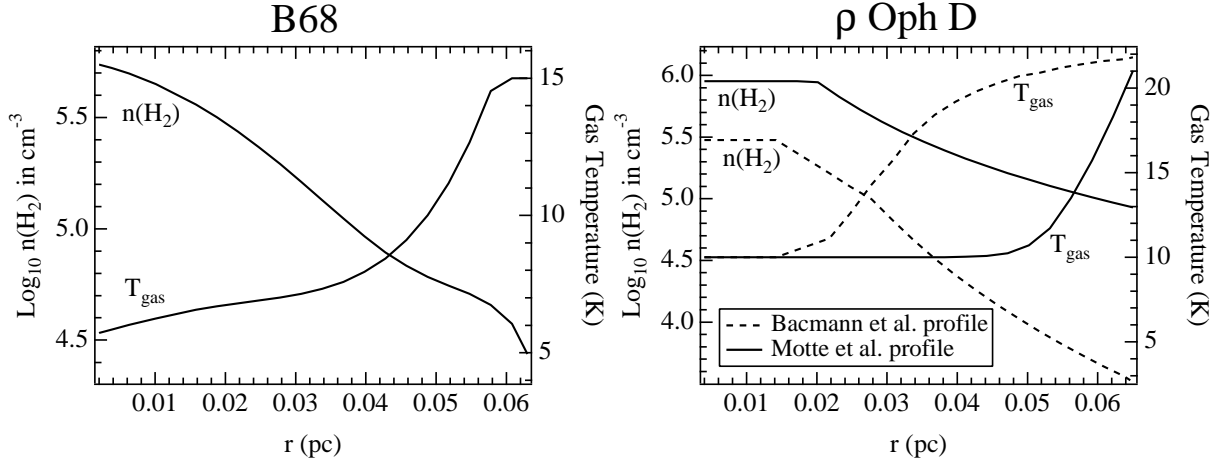}
\caption{Adopted density and temperature profile with radius in B68 (left)
and $\rho$ Oph D (right) molecular cores.  In the latter core we present the density
profile derived by Motte et al. (1998) and Bacmann et al. (2000). }
\end{figure*}

\begin{figure*}
\epsscale{0.65}
\plotone{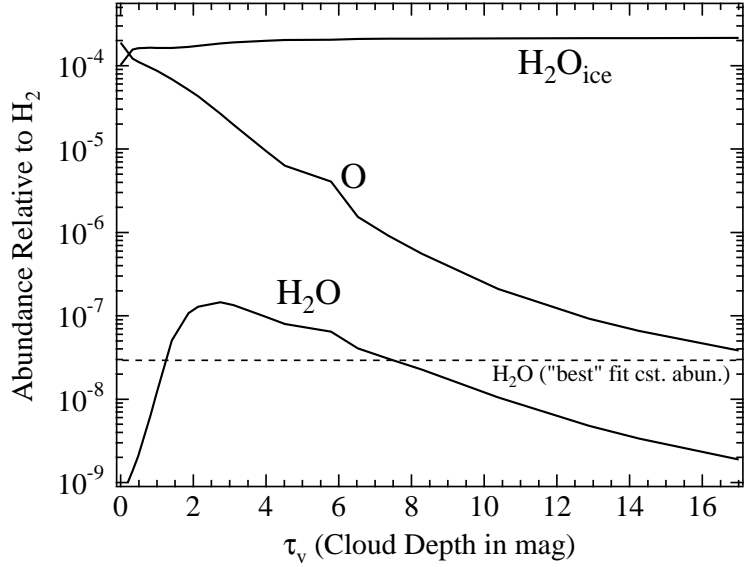}
\caption{Theoretical predictions of molecular abundance as a function of cloud
depth (extinction) from a gas-grain chemical model.  Also shown is the best
fit constant abundance model for the B68 cloud.  
Details of this model are given in \S3.1.}
\end{figure*}

\end{document}